\documentclass[prl,twocolumn,amsmath,amssymb, aps,superscriptaddress]{revtex4}

\usepackage{graphicx}
\usepackage{dcolumn}
\usepackage{bm}
\usepackage{xcolor}

\begin{document}

\title{Liquid-Liquid Phase Separation in an Elastic Network}

\author{Robert W. Style}
\affiliation{%
Department of Materials, ETH Z\"{u}rich, Switzerland.
}%
\email{robert.style@mat.ethz.ch}

\author{Tianqi Sai}
\affiliation{%
Department of Materials, ETH Z\"{u}rich, Switzerland.
}%
\author{Nicol\'{o} Fanelli}
\affiliation{%
Department of Materials, ETH Z\"{u}rich, Switzerland.
}%
\author{Mahdiye Ijavi}
\affiliation{%
Department of Materials, ETH Z\"{u}rich, Switzerland.
}%

\author{Katrina Smith-Mannschott}
\affiliation{%
Department of Materials, ETH Z\"{u}rich, Switzerland.
}%

\author{Qin Xu}
\affiliation{%
Department of Materials, ETH Z\"{u}rich, Switzerland.
}%

\author{Lawrence A. Wilen}
\affiliation{%
Yale University, New Haven, CT 06520, USA.
}%

\author{Eric R. Dufresne}%
\affiliation{%
Department of Materials, ETH Z\"{u}rich, Switzerland.
}%
\email{eric.dufresne@mat.ethz.ch}

\date{\today}


\begin{abstract}
Living and engineered systems rely on the stable coexistence of two interspersed liquid phases. 
Yet surface tension drives their complete separation.  
Here we show that stable droplets of uniform and tuneable size can be produced through arrested phase separation in an elastic matrix.
Starting with an elastic polymer network swollen by a solvent mixture, we change
the temperature or composition to drive demixing.
Droplets nucleate and grow to a stable size that is tuneable by the network cross-linking density, the cooling rate, and the composition of the solvent mixture.
We discuss thermodynamic and mechanical constraints on the process. 
In particular, we show that the threshold for macroscopic phase separation is altered by the elasticity of the polymer network, and we highlight the role of internuclear correlations in determining the droplet size and polydispersity.
This phenomenon has potential applications ranging from colloid synthesis and structural colour to phase separation in biological cells.
\end{abstract}

\pacs{Valid PACS appear here}

\maketitle

Nucleation and growth of liquid droplets is a ubiquitous process.
In the sky above us, it underlies the formation of clouds.
Inside our own cells, the condensation of protein-rich droplets helps to regulate the translation of RNA, among other cellular phenomena \cite{bran11,mitr16}. 
The essential thermodynamics of condensation was worked out in the nineteenth century by Gibbs \cite{gibb06}.
The kinetics of nucleation is more complex and depends sensitively on the presence of impurities \cite{cahn65,lame50,avra39,ande02}. 

A growing appreciation for the biological implications of the phase separation of proteins within living cells \cite{zwic14,hyma14,bran11,pate15,zhu15,moll15,riba17} raises a host of questions about the underlying physics \cite{bran15}.
Examples include the role of active processes in determining the properties of phase-separated droplets \cite{bran11b} and the combination of phase separation with kinetic arrest \cite{dufr09,pate15}.
Of particular interest here is the interplay of the structure of the cytoplasm and phase separation.
The physics of droplet growth has been extensively studied when the surrounding matrix is a simple fluid such as a vapour or Newtonian liquid.  
However a living cell has a complex rheology, including significant elasticity from the cytoskeleton \cite{gard04,wen11}.

Here, we investigate  nucleation and growth of liquid droplets inside of a polymer network.
We show that the resulting droplets are stable and highly uniform, with a size that can be tuned by the cross-linking density,  quench rate, and   loading of the minority fluid. 
Condensation in a compliant elastic network is a generic physical process for making monodisperse droplets, and  works for a wide variety of chemistries.
We demonstrate its efficacy for both temperature- and composition-driven condensation inside of covalently or physically cross-linked polymer networks swollen with silicone or aqueous solvents.
This process may provide a flexible route to the bulk synthesis of monodisperse, polymeric micro- and nano-particles and enable the self-assembly of flexible, structurally-coloured materials \cite{holt97,koll13}.
The interaction of condensation and network elasticity may play a role in the cellular physiology of phase-separated proteins, and the physical parameters identified here could be exploited by living cells to  regulate phase separation \cite{dufr09,feri13}.

\subsection{Liquid-Liquid Phase Separation in an Elastic Network}

\begin{figure}
\centering
\includegraphics[width=1\linewidth]{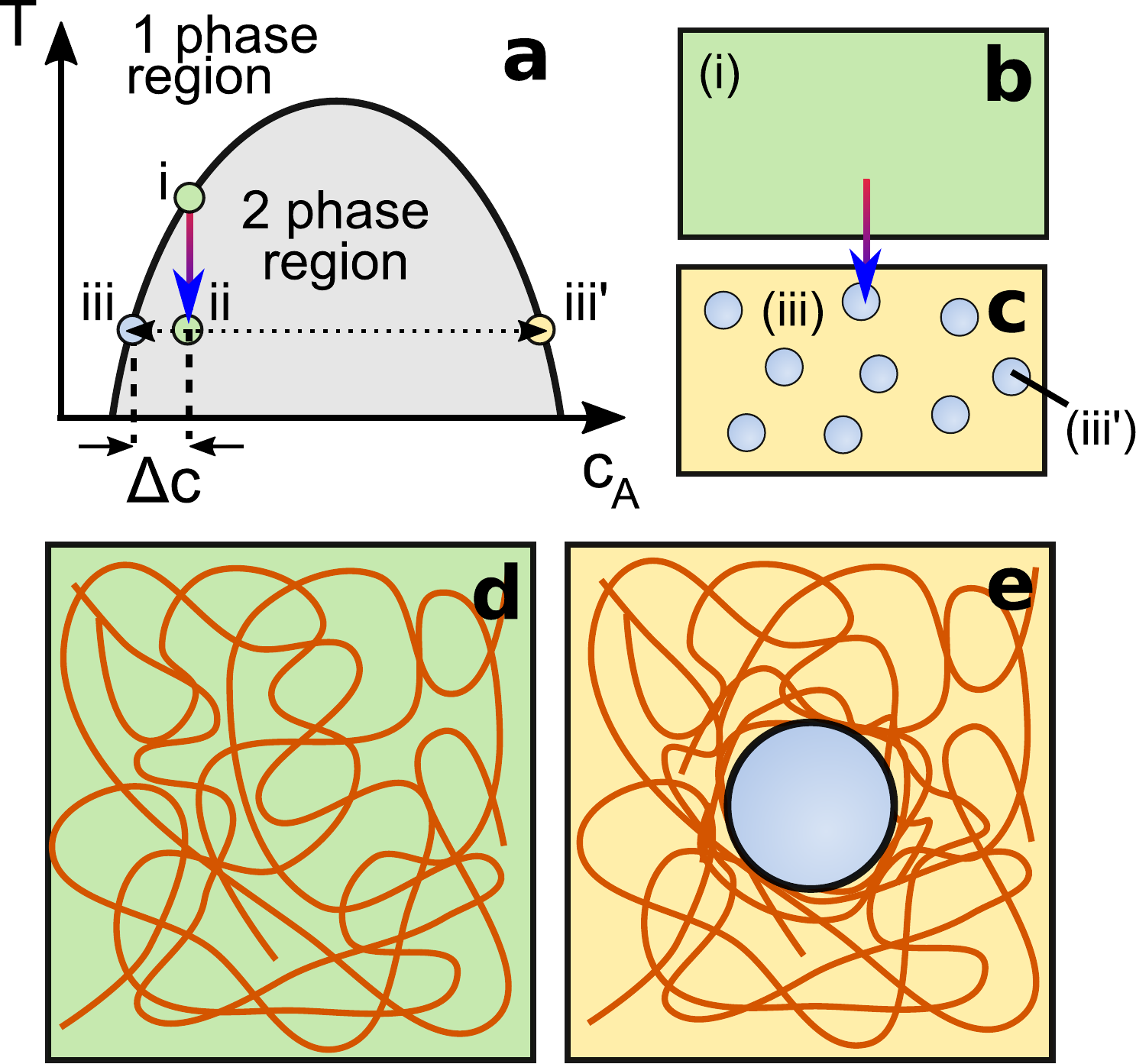}
\caption{Phase separation of a liquid mixture in a polymer network. (a-c) Schematic diagram of simple liquid-liquid phase separation.  Point (i) on the phase diagram in indicates a marginally stable mixture where the majority component, B, is saturated with a dilute component, A. Upon cooling to (ii), the system spontaneously separates into two phases, a continuous phase (iii) with a low concentration of A and a droplet phase rich in A (iii').  (d,e) Schematics of (d) a polymer network swollen by a mixture of A and B (at point (i) in the phase diagram), and (e) the same system after quenching to (ii) and phase separating.  Note that the droplet must deform the network in order to grow beyond the mesh size.
}
\label{fig:schematic}
\end{figure}

The stability of a fluid mixture depends on the temperature and the concentrations of its components (e.g. \cite{cahn65,herz07,hyma14}). 
A  schematic phase diagram for a typical mixture of two liquids, A and B, is shown in Figure \ref{fig:schematic}a.  
Above a critical temperature, any mixture of the two fluids is stable. 
Below the critical temperature, some mixtures -- indicated by the gray region in Figure \ref{fig:schematic}a -- are unstable.
Now consider a marginally stable fluid mixture at  the boundary between the stable and unstable regions, as indicated by point \emph{(i)}.
When the mixture is cooled rapidly to point \emph{(ii)} in the unstable region, \emph{i.e.} `quenched', it spontaneously separates into two stable compositions (Figure \ref{fig:schematic}b,c) . 
When the quench is not too deep,  separation happens through the nucleation of discrete droplets rich in A \emph{(iii')}, which grow until the concentration of A reaches its saturation value in the continuous phase \emph{(iii)}.
Since the interface between the droplet and the continuous phase costs energy, the droplets tend to coarsen over time.  
Just like forced mixtures of oil and water, droplets continue to grow until  there is a single blob of each phase.
This can occur by direct coalescence or through diffusion of the minority component through the continuous phase (\emph{i.e.} Ostwald Ripening).  

We investigate how this classic process is modified by the presence of a polymer network.
Consider a polymer network that is swollen by a mixture of two solvents, A and B, as shown in Figure \ref{fig:schematic}d.
Imagine that this solvent mixture initially lies at point \emph{(i)} on the phase diagram and is quenched to the unstable point \emph{(ii)}. 
As before, the fluid mixture needs to separate.
For simplicity, let's assume that the polymer network is excluded by the A-rich droplet phase.
Provided that the network is not too dense, initial nucleation and growth will not be affected by the network.
However once the droplet is comparable or bigger than the network mesh size, it cannot grow without deforming the network (Figure \ref{fig:schematic}e).
How does this impact  phase separation?  
At a minimum, we expect the network to prevent motion of the droplets, blocking direct coalescence as a route to coarsening.
More intriguingly, the thermodynamic forces which drive condensation could be balanced by elastic forces in the polymer network, arresting growth.

When a mixture is quenched into the two-phase regime, it is imbued with excess free energy that can deform a polymer network. 
Consider an unconstrained liquid in equilibrium with a dilute solute. Here the solute concentration, $c$, will take its saturation value, $c_{sat}(T)$.
If $c$ is increased beyond $c_{sat}$,  the solute will condense into the liquid phase until equilibrium is achieved.
This is driven by an excess free energy per solute molecule  given by $\Delta g(T,c) = k_BT \ln (c/c_{sat})$, where $k_B$ is Boltzmann's constant. 
Here we have assumed the solute is dilute enough to be ideal.  
However if the liquid phase is constrained by the elastic network of a gel, it can have an increased internal pressure, $P$.
Now the pressure and volume in the droplet will increase as the solute condenses until the work done by the condensing solute to grow the droplet, $P\Delta V=PM/\rho_lN_A$, balances $\Delta g$ (here $\rho_l$ is the density of the liquid, $M$ is the molar mass of the solute, and $N_A$ is Avogadro's number). Thus, in equilibrium, we expect
\begin{equation}
P = \frac{\rho_{\ell} RT}{M} \ln \left(\frac{c}{c_{sat}}\right),
\label{eqn:p}
\end{equation}
where $R=k_B N_A$ is the gas constant (\emph{cf} \cite{whee08,cai10}).
If the restraining force of the polymer network exceeds this pressure, it cannot grow.
This is perfectly analogous to the stalling of a processive molecular motor by a sufficiently large opposing force \cite{schn00}.
Note that these arguments do not rely on phase-separation being driven by changes in temperature.
Thus Equation (\ref{eqn:p}) can also  be used to estimate the driving pressure when phase separation is triggered  by other processes, such as changes in solvent or solute composition.

Droplets grow freely when they are smaller than the network mesh size, but to grow beyond the mesh size, they have to deform the network.
Applying the classic theory of elastic cavitation \cite{gent59,ball82,gent74,gent91,zimb07,kund09} to droplet growth, we expect two regimes delimited by a critical pressure $P_{\mathrm{crit}} = 5E/6$, where $E$ is Young's modulus of the gel.
When $P<P_{\mathrm{crit}}$, a growing droplet should make only modest deformations to the network, and will not grow much beyond the network mesh size.
However when $P>P_{\mathrm{crit}}$, droplet growth cannot be stopped by a linear-elastic (or Neo-Hookean) material (\emph{e.g.} \cite{gent59,zimb07,zhu11}), and is only limited by the availability of solute.
Equivalently, since elastic forces make nucleation and growth energetically unfavourable at small supersaturations, the apparent phase boundary for macroscopic phase separation is shifted:
\begin{equation}
c_{sat}^{app}=c_{sat}(T)e^{\frac{5EM}{6\rho_l R T}}.
\end{equation}

Simple scaling arguments suggest that the condition for solute-limited growth can readily be satisfied in most solvent-swollen polymer networks.  
Ideal rubber elasticity theory \cite{gent74} relates Young's modulus to the structure of the polymer network,  $E \approx n_c kT$.  
Here, $n_c$ is the number density of crosslinks in the network.
Combining this result with Equation (\ref{eqn:p}), we find $P/E \approx n_l/n_c$, where $n_l$ is the number density of molecules in the droplet phase  (assuming that $\ln(c/c_{sat})\sim O(1)$).
Thus, we find that droplets can strongly deform a polymer network whenever $n_l\gtrsim n_c$, or equivalently when the size of a molecule in the  droplet is smaller than the  mesh size of the polymer network.
Polymer networks swollen with a reasonably supersaturated mixture readily meet this condition.

\begin{figure*}
\centering
\includegraphics[width=1\linewidth]{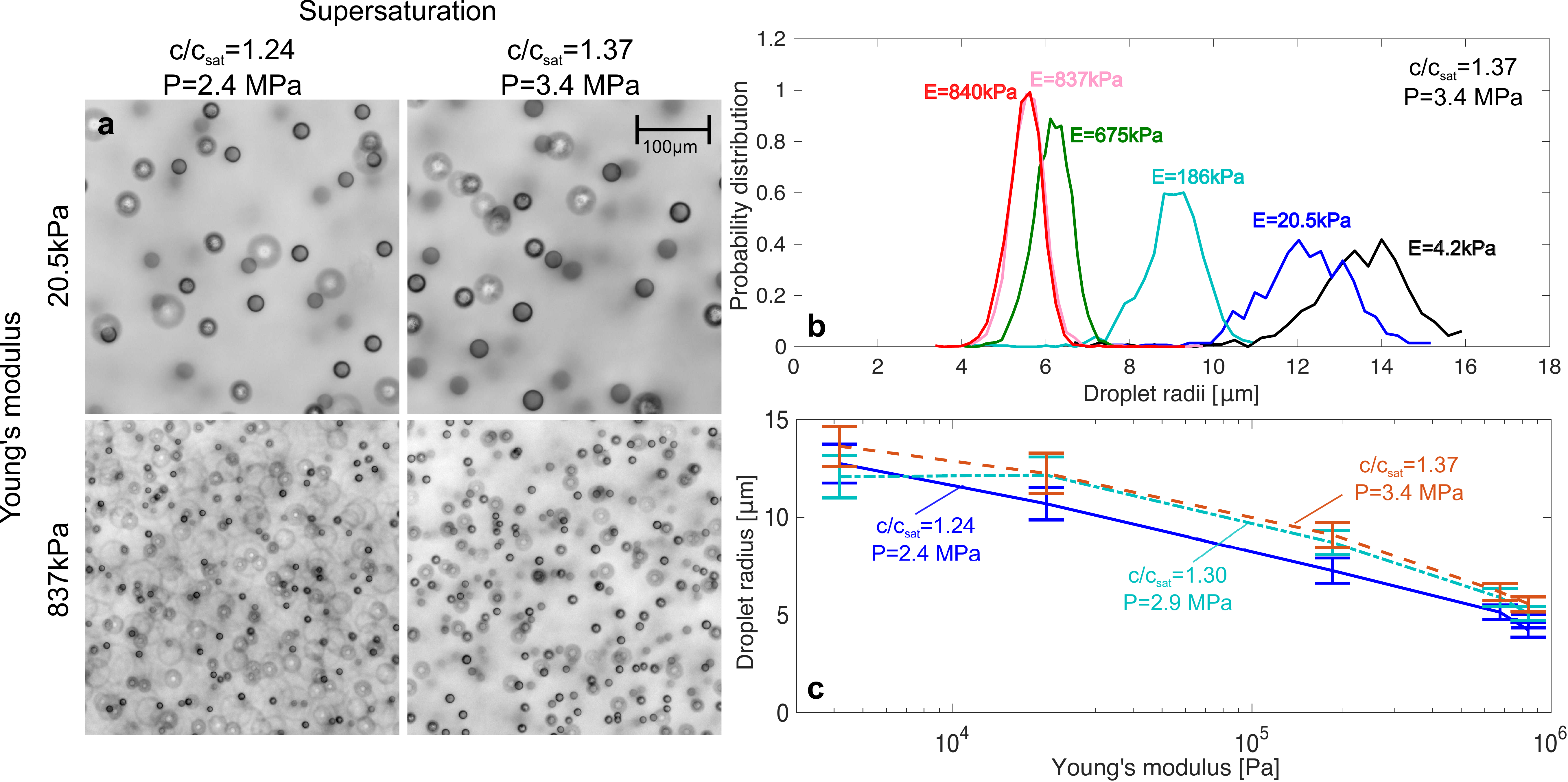}
\caption{Droplets formed by phase separation of fluorinated oil in silicone gels. a) Examples of droplets formed in a soft (top), and stiff (bottom) gels. Examples in the left column have a lower loading than the right column, as they are saturated at 37/44$^\circ$C respectively, before being cooled at a rate of 1.5$^\circ$C/min to 23$^\circ$C. b) Examples of typical droplet distributions in different stiffness gels, each saturated at 44$^\circ$C. c) The dependence of  droplet size on stiffness and saturation temperature. Data points are centred on the average value and the error bars indicate the standard deviation. In each panel, $P$ is calculated using Equation (\ref{eqn:p})}
\label{fig:droplets_in_silicone}
\end{figure*}

\subsection{Thermally-induced phase separation in silicone gels}
To demonstrate this process, we grow droplets of fluorinated oil in a silicone gel.
The silicone gel consists of a cross-linked silicone polymer network swollen in silicone oil.
By varying the cross-linking density, we can tune Young's modulus, $E$, from about 5 kPa to 1 MPa (\emph{e.g.} \cite{styl15}).
We saturate the gels with fluorinated oil at an elevated temperature, $T_e$.  
The saturation concentration of the oil in the gel has no significant dependence on the cross-linking density, but increases with temperature in the range 23 - 55 $^\circ$C, as $c_{\mathrm{sat}}[\mathrm{wt\%}]=0.093 \, T_e [^\circ \mathrm{C}]+3.2$ (see Supplemental Material).
To initiate droplet formation, we cool samples to 23$^\circ$C at a controlled rate.

We  estimate the pressure available to to deform the elastic network using Equation (\ref{eqn:p}).
For fluorinated oil ($\rho_l/M=4494\,\mathrm{mol}/\mathrm{m}^3$), the pressure pre-factor $\rho_l RT/M = 11 \mathrm{ MPa}$.
In these experiments, we can readily reach $c/c_{sat}$ up to 2, corresponding to a driving pressure of 7.7 MPa.
This exceeds the anticipated threshold for  droplet growth, even for the stiffest gels we consider here.

\subsubsection{Tuning Droplet Size}
Over a wide range of experimental conditions we observed uniform micron-scale spherical droplets distributed homogeneously throughout the sample.
Typical microstructures of the resulting droplet dispersions are shown in Figure \ref{fig:droplets_in_silicone}a, and bright-field optical $z$-stacks are found in \textbf{Supplemental Videos 1-4}.
The droplets are fixed in position and show no visible changes in their radii over  time scales of hours.  
At longer time scales, the droplets  shrink as the oil evaporates.
In contrast, droplets of fluorinated oil nucleating and growing in liquid silicone coarsen to the millimetre scale within seconds of cooling. As described below, the size  of the droplets depends on the cross-linking density, the level of saturation, and the quench rate.

The droplet size is primarily controlled by the cross-linking density of the gel.
Analysis of the bright-field $z$-stacks allows us to size each droplet and construct size distributions, as shown in Figure \ref{fig:droplets_in_silicone}b.
In these data, samples are saturated at T=44$^\circ$C and quenched to 23$^\circ$C at  1.5$^\circ \mathrm{C}$/min.  
This corresponds to a driving pressure of 3.4MPa.
We varied the cross-linking density to achieve Young's moduli from 4.2 to 840kPa. 
In all cases, the droplets have single-peaked size distributions that are very well approximated by normal distributions with mean $\mu$ and standard deviation $\sigma$.
The mean droplet radius varies from about 5 $\mu$m  to 14 $\mu$m.  
The decrease in mean droplet size with gel stiffness is shown in Figure \ref{fig:droplets_in_silicone}c. 
The dependence on Young's modulus is modest, such that a 200-fold increase in gel stiffness only decreases the droplet size by a factor of 2.

Droplet size is also impacted by the level of saturation. 
By varying the  incubation temperature, we  varied the loading of the oil from $c/c_{\mathrm{sat}}=$1.24 to 1.37 wt\%.
This 10\% increase in loading (corresponding to a 50\% increase in the supersaturation, $c-c_\mathrm{{sat}}$) leads to a 25\% increase in droplet radius, as shown in Figure \ref{fig:droplets_in_silicone}c.

\begin{figure}
\centering
\includegraphics[width=1\linewidth]{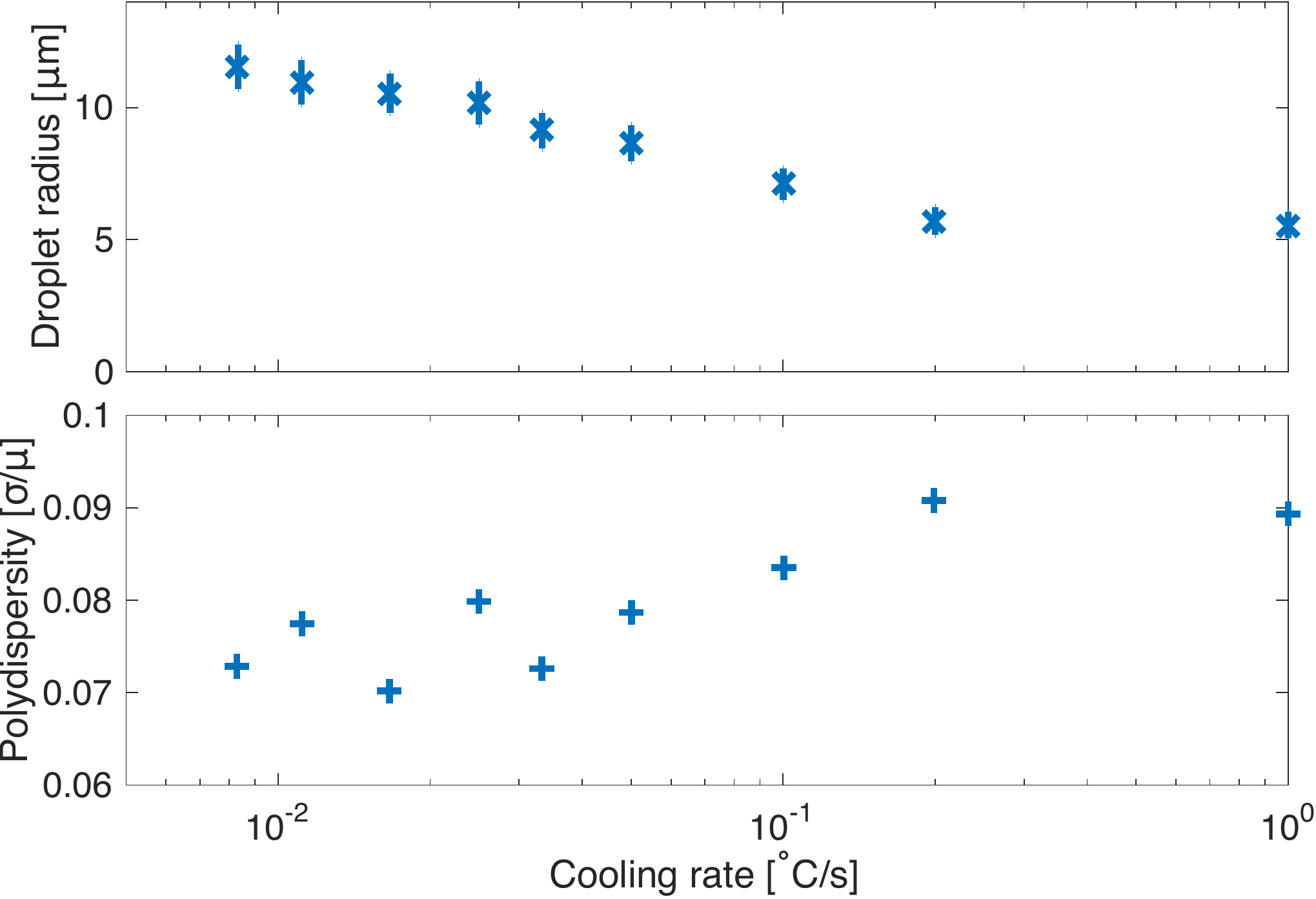}
\caption{Rate dependence of droplet size and polydispersity. (top) The mean droplet radius reduces roughly logarithmically with the cooling rate.  Here, the error bars are the standard deviation, $\sigma$, of the droplet size distribution.  (bottom) The polydispersity ($\sigma/\mu$) is effectively independent of the cooling rate. $E=186$kPa, and the sample is saturated at $40^\circ$C, so that $c/c_{\mathrm{sat}}=1.37$.}
\label{fig:cooling_rate}
\end{figure}

We can also control droplet size by controlling the cooling rate:
droplets decrease in size as the cooling rate increases, 
as shown in Figure \ref{fig:cooling_rate}a.
Again, we find a modest, roughly logarithmic, dependence of droplet radius of cooling rate.
Droplets are slightly more uniform when cooled at a slower rate.
We quantify the uniformity with the polydispersity index ($\sigma/\mu$), shown in Figure \ref{fig:cooling_rate}b.
Over the range of two decades in cooling rate, the polydispersity ranged from 7 to 9 \%.
Thus, we can control the size this way with surprisingly little penalty in droplet uniformity as the cooling rate increases.

\subsubsection{Optical Properties of Composites with Uniform Droplets}

\begin{figure}
\centering
\includegraphics[width=1\linewidth]{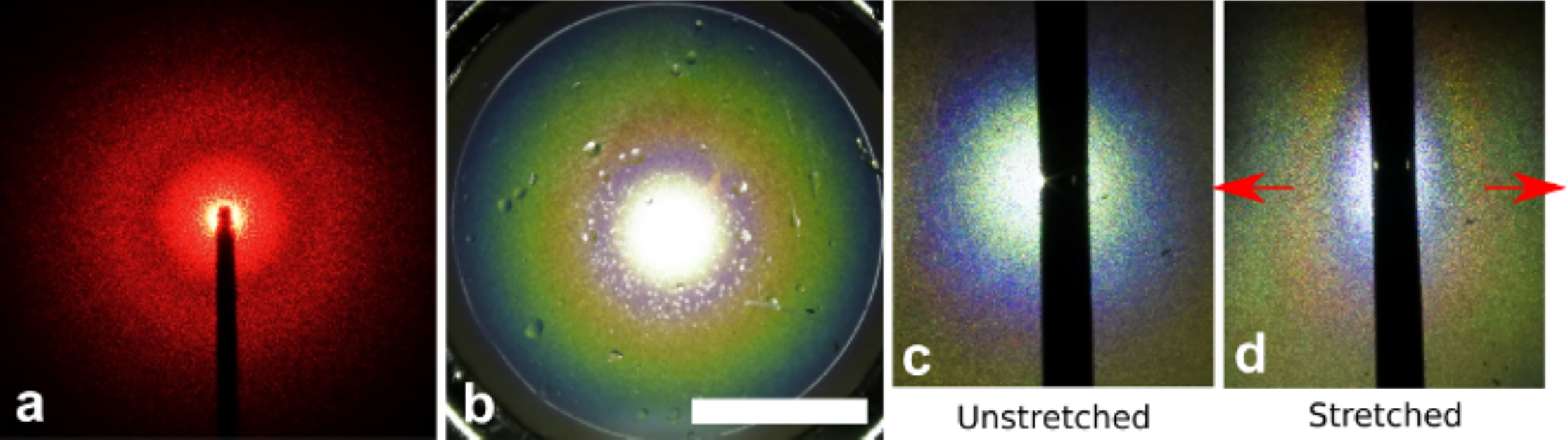}
\caption{Optical properties of self-assembled composites of uniform fluorinated-oil droplets in a silicone gel matrix. (a) The light-scattering pattern of a Helium-Neon laser by a sample ($E=837$kPa silicone gel saturated with fluorinated oil at 40$^\circ$C and cooled at 30$^\circ$C/min)  has a clear ring, indicating droplet uniformity. The centre of the dark ring is at a scattering angle of 6.5$^\circ$. (b) Different wavelengths of light are scattered to different angles, leading to colourful rings when a sample (Sylgard 184, $E=40$kPa, cooled from 42$^\circ$C at 6$^\circ$C/min) is illuminated from behind with a white-light LED. The scalebar is 10mm long. (c,d) The color pattern is deformed upon stretching (Sylgard 184, $E=412$kPa, cooled from 42$^\circ$C at 6$^\circ$C/min). Here, the direct light from the white-light LED source is blocked by a 2mm-thick rod to better visualise the colours. See Supplemental Video 2.}
\label{fig:light}
\end{figure}

The uniformity of the droplet size is reflected in the macroscopic optical properties of the gel.
Figure \ref{fig:light}a shows the scattering pattern in transmission for a thin section of silicone gel containing fluorinated-oil droplets, illuminated with a HeNe laser.
There is a clear ring, consistent with the expected Mie scatting pattern for monodisperse spheres \cite{bohr08}.
There is no significant contribution to the observed scattering pattern from correlations in the droplet locations (\emph{i.e.} structure factor).
Since the silicone gels are highly elastic, we can deform the droplets by stretching the gel \cite{styl15}.
As shown in \textbf{Supplemental Video 5}, this yields a corresponding stretching of the scattering pattern in the direction perpendicular to the applied stretch -- indicating the coupling of the macroscopic deformation to the  shape of the microscopic droplets.
Different wavelengths are scattered to rings of different radii by the samples, so we observe a pattern of color when a point source of white light is viewed through the sample (Figure \ref{fig:light}b).
This pattern  changes as the sample is stretched and the microstructure is deformed, as shown in Figure \ref{fig:light}c,d and \textbf{Supplemental Video 6}.

\subsubsection{Correlated Nucleation and Growth}

As expected for gentle quenches far from the critical point, droplets form by subsequent nucleation and growth  (\emph{e.g.} \cite{mery56,lepe14}).
\textbf{Supplemental video 7} shows a typical example.  Here  $E=186$kPa and the  sample is cooled from 42$^\circ$C to 22$^\circ$C at 2$^\circ$C/min.
New droplet nuclei appear over a period of about 150s, and grow over a longer interval, of about 800s.
This suggests that nucleation and growth can be thought of as two separate stages of droplet formation \cite{lame50}.
Here we show that while these two processes are separated in time, they are strongly coupled in a manner than impacts the droplet size distributions.

\begin{figure}
\centering
\includegraphics[width=1\linewidth]{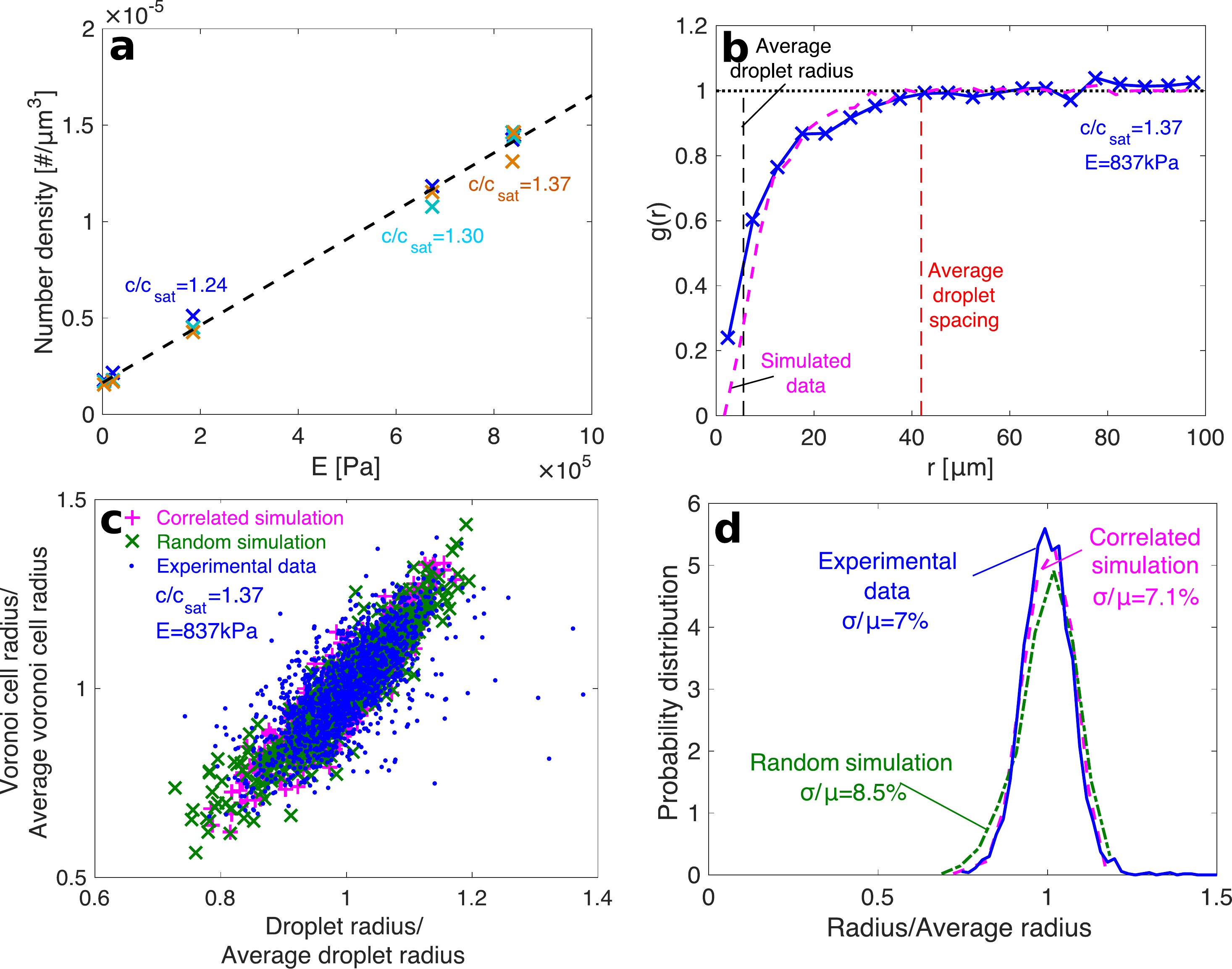}
\caption{Correlated nucleation and growth.  a) The number density of droplets depends on stiffness, but not saturation. The dashed line is the line of best fit to the data.
b) The pair-correlation function, $g(r)$ of droplet centres for a typical experiment, blue crosses, shows that nuclei do not form near each other. The range of internuclear repulsions is comparable to the inter-droplet distance. The pink line shows the $g(r)$ for nuclear positions in our Monte-Carlo growth simulation. Note that the experimental measurements of $g(r)$ are only accurate for $r\gtrsim 5\mu$m, due to  inaccuracy of out-of-plane tracking of droplet centres.
c) Droplet sizes are strongly correlated with the radius of their voronoi cell. Blue: data from the experiment in b). Green and pink points show results of growth simulations with random and correlated nuclear positions.
d) Droplet size distributions for the experimental data in b) (blue) and simulations with random (green) or correlated (pink) positions.}
\label{fig:monodispersity_mech}
\end{figure}

Since droplets are trapped in an elastic matrix,  nucleation positions are given by the final positions of the droplets.
We identified the position of each nucleus using the same bright-field $z$-stacks underlying Figure \ref{fig:droplets_in_silicone}.
First, we determined the number density of droplets, $n_d$, as a function of cross-linking density and supersaturation.
As shown in Figure \ref{fig:monodispersity_mech}a, $n_d$ increases linearly with Young's modulus, but is independent of the supersaturation.
By a simple application of mass conservation, this variation in nucleation number density fully accounts for the dependence of the mean droplet size on the cross-linking density and extent of supersaturation (see Supplement).

In classical theory the interaction of nucleation events is typically ignored. However we find that the locations of nucleation sites are significantly correlated, and central to the form of the final droplet distribution. 
We quantified spatial correlations in the nucleation positions using the radial distribution function, $g(r)$, as shown in
Figure \ref{fig:monodispersity_mech}b for typical experimental data.
The pair-correlation function compares the  probability of observing two objects at a given separation, relative to a case where all objects are placed perfectly randomly.  
We find that the number density of closely-neighbouring droplets is strongly reduced over a distance comparable to the mean inter-droplet spacing,  about 40$\mu m$ in the example of Figure \ref{fig:monodispersity_mech}.
This suggests that there is some mechanism that prevents droplets from nucleating near each other.

Intriguingly, variations in the local spacing of nuclei have a significant impact on the size distribution.  
We  characterised the structure about each nucleus using the Voronoi construction.
The Voronoi cell associated with a nucleus is the collection of points that are closer to it than any other nucleus. 
In general, this is a complex polyhedral shape.  
We reduce it to a single length scale, by taking the cube-root of the cell volume, or Voronoi radius, $r_v$.
As shown in Figure \ref{fig:monodispersity_mech}c, the Voronoi radius is strongly correlated to the droplet size.
Thus droplets in tight clusters tend to be smaller than more widely-spaced droplets.
This effect, if present in conventional liquid-liquid phase separation, is masked by Brownian motion  and coalescence of droplets.

Monte-Carlo simulations suggest that the correlation of droplet size and nuclear spacing is a natural consequence of the diffusion-limited growth of droplets.
Using Brownian dynamics, we simulated the diffusion and capture of solute molecules by a number of fixed nuclei.
In one case, the nuclei were positioned completely randomly, in the other case, the nuclei positions were selected  to have the same $g(r)$ as experiments.
In both cases, we find a pronounced correlation between $r_v$ and droplet radii, as shown by the green (random) and pink (correlated) points in Figure \ref{fig:monodispersity_mech}c, which nicely match the experimentally-observed correlation (blue).
However, the simulation with correlated nucleation positions shows a narrower spread in $r_v$ and droplet radii, which closely matches the experiments, as shown in Figure \ref{fig:monodispersity_mech}d.
Thus long-range interactions of nuclei make the structure around each nucleus more regular and result in more uniform droplet radii.

Together, these results suggest that understanding the nucleation process is crucial to gaining a quantitative understanding of droplet size and monodispersity.

\subsection*{Isothermal Phase Separation in Hydrogels}

To show the generality of uniform droplet production by phase separation in a polymer network, we demonstrate its efficacy for a distinct form of phase separation and diverse polymeric networks.

Previously we supersaturated the system by quenching a binary liquid mixture with a change of temperature.
Now we drive phase separation of a solute by changing the composition of a solvent mixture \cite{lame50,vita03}. 
This form of phase separation is familiar from aperitifs like ouzo and pastis.
In these drinks, fragrant anise oil is solubilised in water by a high concentration of ethanol \cite{tan16}.
When water is added to the drink, the mixture becomes unstable,  and oil-rich droplets nucleate and grow.

\begin{figure}
\centering
\includegraphics[width=1\linewidth]{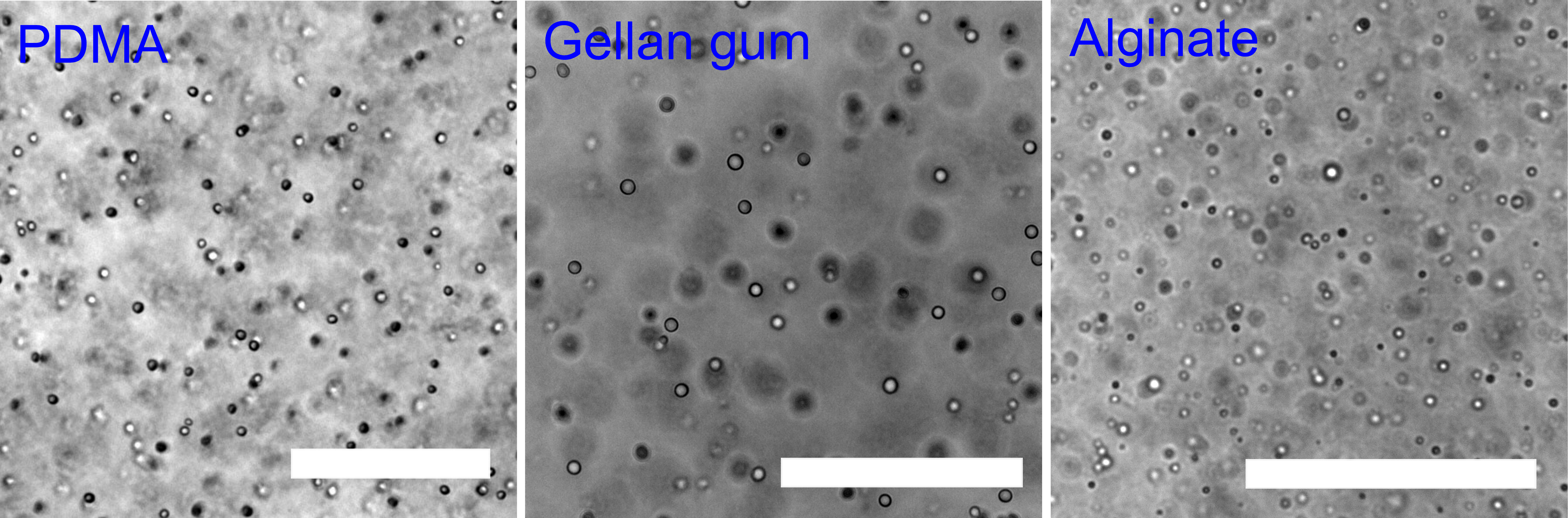}
\caption{Arrested isothermal phase-separation in diverse hydrogels.  Here, anise oil droplets nucleated and grew in an ethanol-water mixture (\emph{i.e.} the Ouzo effect).  Micrographs of dispersed droplets in (a) PDMA, a covalently crosslinked gel, (b) gellan gum, a thermally set physically crosslinked gel, and (c) alginate, an ionically crosslinked physical gel. All droplets are stable against coalescence.  Droplets in PDMA and gellan are reasonably monodisperse (see Supplement). The scale bars are all 40$\mu$m wide.}
\label{fig:hydrogels}
\end{figure}

We use three hydrogels with distinct chemistries: a chemically-crosslinked gel (polydimethylacrylamide or PDMA), a thermally-set physical gel (gellan gum), and an ionically-crosslinked physical gel (alginate).  
In each case we soak the hydrogel in a large volume of a stable mixture of water/ethanol/anise oil to exchange the solvent (see Figure \ref{fig:schematic}b).
After this we cover the gel with deionized water, which diffuses inward and initiates phase separation.
The resulting oil droplets are shown in Figure \ref{fig:hydrogels}.
As with the fluorinated-oil/silicone system, we see rather uniform droplet distributions in the different materials (see Supplement for distributions), especially in the PDMA and gellan gum samples which we expect to be much more uniform gels than alginate \cite{hein05}.

\subsection*{Conclusions}
We have shown that phase separation of a solvent mixture in a polymer network is a simple bulk technique to create uniform droplets.
The size of the resulting droplets depends on a range of factors including the cross-linking density of the polymer network, the supersaturation, and the quench rate. 
This process for the production of monodisperse droplets in bulk has broad potential applications.
For example, it could be used to create composites with well-defined microstructures that enable novel optical or mechanical properties \cite{dufr09,styl15}.
Additionally, by using a monomer as the phase-separating component, it can offer a flexible route for the synthesis of uniform microparticles.

Our results suggest a role for polymer networks in the regulation of membrane-less organelles within living cells.
Consider phase-separating macromolecules with molecular weights in the range of 75-600 kDa.
Equation (\ref{eqn:p}) suggests they can exert pressures of order $4-33$kPa.
These pressures are comparable to the stiffness of many cytoskeletal networks, which suggests that living cells may be able to regulate the nucleation and growth of membrane-less organelles through their mechanical properties.

While we have outlined the essential phenomenology and highlighted key thermodynamic and mechanical aspects of the process, much work must be done to yield a quantitative understanding. 
The central question is how an increase in the cross-linking density leads to an increase of nucleation sites. 
There are two key aspects to nucleation -- its spatial distribution, and rate.
The spatial distribution is likely to be set by the fact that when a droplet nucleates and starts to grow, it will deplete the surrounding material of solute, making it less likely for a nearby nucleus to appear.
The control of nucleation rate is more complex, as it will depend on a range of factors, such as cooling rate and the ability of droplets to nucleate within the (nanometric-scale) mesh of the polymer gel.
However the monotonic increase of nucleation sites with cross-linking density suggests the simple possibility that the cross-linking sites themselves serve as heterogeneous nucleation sites. 

The mechanical aspects of this process demand further exploration.
We have shown that the mechanical properties of the gel alters the phase diagram. 
However there are number of further interesting questions and possibilities.
While elastic cavitation theory predicts a balance of elastic forces and condensation pressure only for droplets that are modestly larger than the mesh size, it is clear that strain-stiffening networks can suppress droplet growth and limit droplet size over a wider range of driving pressures.
Furthermore, elastic forces could create a elastic Ostwald ripening effect. 
In classic Ostwald ripening, surface tension favors the collapse of small drops to feed big ones.  
Here, we anticipate that elastic forces in strain-stiffening networks could drive transport of the condensed phase from big drops to small ones.
The balance of these two effects could lead to a mechanically defined equilibrium droplet size. 
However precise quantitative theories of these effects will be challenging, since the deformations are so large -- here, the final droplets are about three orders of magnitude larger than the network mesh size. 
Therefore there must be strong irreversible deformations -- especially in the vicinity of the droplet -- which are bound to be strongly rate-dependent \cite{gent91,macm15,hutc16,cret16}.

\subsection{Materials \& Methods}

\subsection*{Silicone gels preparation and characterisation}
We prepare silicone gels following the recipe of \cite{styl15}, except where stated otherwise. 
Stiffness is tuned by varying the proportion of crosslinker in the gel between 0.8\% and 3.3\%.
We measure the Young's modulus of each sample by making an additional bulk sample and performing a compression test.
For this, the silicone gel is formed in a cylindrical plastic mould (depth 10mm, radius 15mm), and then indented with a 1mm radius, cylindrical indenter  using a TA.XT plus Texture Analyser with a 500g load cell (Stable Micro Systems).

For some of the light-scattering experiments, we used a different variety of silicone gel (Sylgard 184, Dow Corning), as this is less sticky, and thus easier to stretch manually.
For these samples, stiffness was again tuned by varying the ratio of base to crosslinker.

Silicone-gel samples are made by coating the inside of glass-bottomed petri dishes with a thin, $O(1$mm$)$-thick film of silicone gel.
Silicone gel samples are then saturated with fluorinated oil (Fluorinert FC770, Fluorochem) at an elevated temperature, $T_e$.

The solubility of fluorinated oil in the gels was measured by weighing samples of silicone gels of two different stiffnesses ($E=20,840$kPa) before and after saturation at 23,40,55$^\circ$C. No significant change in solubility was found with $E$. See Supplemental Material for further details.

\subsection*{Growth simulations}

We explored the effect of inter-nuclei structural correlations on the growth of droplets using Monte-Carlo simulations.  In the first step, we positioned $10^3$ nuclei in a cube with periodic boundary conditions, either by placing the nuclei independently or using the Metropolis algorithm to generate a pattern of points with the same $g(r)$ as observed in experiments. In the second step, we released $10^6$ molecules, which underwent a random walk until they were captured by a nucleus.  By counting how many molecules were captured by each nucleus, we were able to determine the final size of each nucleus. For simplicity, the target radius for each nucleus was fixed throughout the simulation so that the volume fraction of the targets was 1.25\%.

\subsection*{Hydrogel preparation}

PDMA samples were fabricated by free-radical polymerisation of N-N-dimethylacrylamide (DMA, Sigma-Aldrich) with N,N'-methylenebisacrylamide (MBA, Sigma-Aldrich) as a crosslinker. Ammonium persulphate (APS, Fisher-Bio) and N,N,N',N'-tetramethylethylenediamine (TEMED, Apollo Scientific) were used as the redox/initiator system (\emph{e.g.} \cite{orak07}). All chemicals were used as received, and the reaction was performed in a nitrogen glove box to avoid inhibition of the reaction by oxygen.
We prepared fresh stock solutions with deionized water as follows: 30\%w/v DMA, 2\%w/v MBA, 10\%w/v APS and 10\%w/v TEMED.
To make 10mls of PDMA, we combined 2.5ml DMA solution, 0.8ml MBA solution and 6.64ml deionized water before gently mixing.
Then we added 0.03ml TEMED solution, and 0.03ml APS solution, mixing the sample gently after each addition. Finally the sample was poured into a mould and left to completely cure.

Gellan-gum samples were fabricated from 0.5\,wt\% gellan gum (Apollo Scientific Ltd.), 59.5\,wt\% deionized water, and 40\,wt\% ethanol.
The gum powder is mixed with the water, heated to 90$^\circ\mathrm{C}$ and stirred until fully dissolved.
Ethanol is then added drop-wise while stirring and the sample is removed from the heat.
Typically we see a separation of the sample into an upper, cloudy part, and a lower, transparent part.
We pipette the lower, clear phase into petri dishes and allow it to gel.
Finally, the gel is soaked in CaCl$_2$ solution (50-500mM) to allow full crosslinking by the calcium ions.

Alginate-gel samples were fabricated from 2\,wt\% sodium alginate (Acros) in deionized water.
This is gently placed in contact with 100mM CaCl$_2$ solution and left for long enough to allow the calcium ions to diffuse into, and crosslink the alginate solution (\emph{e.g.} \cite{jodd16}).

\subsection*{Further details} Further information about the controlled cooling apparatus, image analysis, and optical experiments are given in the Supplemental Material.

We acknowledge helpful discussions with Alain Goriely and Ian Griffiths and funding from the Swiss National Science Foundation (grants 200021\_172824 and 200021\_172827).

\subsection*{Author contributions} R.W.S., L. A. W. and E. R. D. designed experiments. R.W.S., T. S., N. F., M. I., K. S.-M. and Q. X. performed experiments. R.W.S., L.A.W. and E.R.D analysed data. R.W.S, L.A.W. and E.R.D. developed the theory. R.W.S, and E.R.D. wrote the paper.

\subsection*{Competing financial interests} The authors declare no competing financial interests.


\begin{thebibliography}{10}

\bibitem{ande02}
V.~J. Anderson and H.~N.~W. Lekkerkerker.
\newblock Insights into phase transition kinetics from colloid science.
\newblock {\em Nature}, 416(6883):811--815, 2002.

\bibitem{avra39}
M.~Avrami.
\newblock Kinetics of phase change i. general theory.
\newblock {\em J. Chem. Phys.}, 7:1103, 1939.

\bibitem{ball82}
J.~M. Ball.
\newblock Discontinuous equilibrium solutions and cavitation in nonlinear
  elasticity.
\newblock {\em Phil. Trans. Royal Soc. London A}, 306(1496):557--611, 1982.

\bibitem{bohr08}
Craig~F Bohren and Donald~R Huffman.
\newblock {\em Absorption and scattering of light by small particles}.
\newblock John Wiley \& Sons, 2008.

\bibitem{bran11}
C.~P. Brangwynne.
\newblock Soft active aggregates: mechanics, dynamics and self-assembly of
  liquid-like intracellular protein bodies.
\newblock {\em Soft Matter}, 7(7):3052--3059, 2011.

\bibitem{bran11b}
C.~P. Brangwynne, T.~J. Mitchison, and A.~A. Hyman.
\newblock Active liquid-like behavior of nucleoli determines their size and
  shape in xenopus laevis oocytes.
\newblock {\em Proc. Nat. Acad. Sci.}, 108(11):4334--4339, 2011.

\bibitem{bran15}
C.~P. Brangwynne, P.~Tompa, and R.~V. Pappu.
\newblock Polymer physics of intracellular phase transitions.
\newblock {\em Nature Phys.}, 11(11):899, 2015.

\bibitem{cahn65}
J.~W. Cahn.
\newblock Phase separation by spinodal decomposition in isotropic systems.
\newblock {\em J. Chem. Phys.}, 42(1):93--99, 1965.

\bibitem{cai10}
S.~Cai, K.~Bertoldi, H.~Wang, and Z.~Suo.
\newblock Osmotic collapse of a void in an elastomer: breathing, buckling and
  creasing.
\newblock {\em Soft Matter}, 6(22):5770--5777, 2010.

\bibitem{cret16}
C.~Creton and M.~Ciccotti.
\newblock Fracture and adhesion of soft materials: a review.
\newblock {\em Rep. Prog. Phys.}, 79(4):046601, 2016.

\bibitem{dufr09}
E.~R. Dufresne, H.~Noh, V.~Saranathan, S.~G.~J. Mochrie, H.~Cao, and R.~O.
  Prum.
\newblock Self-assembly of amorphous biophotonic nanostructures by phase
  separation.
\newblock {\em Soft Matter}, 5(9):1792--1795, 2009.

\bibitem{feri13}
M.~Feric and C.~P. Brangwynne.
\newblock A nuclear f-actin scaffold stabilizes rnp droplets against gravity in
  large cells.
\newblock {\em Nature Cell Bio.}, 15(10):1253, 2013.

\bibitem{gard04}
M.~L. Gardel, J.~H. Shin, F.~C. MacKintosh, L.~Mahadevan, P.~Matsudaira, and
  D.~A. Weitz.
\newblock Elastic behavior of cross-linked and bundled actin networks.
\newblock {\em Science}, 304(5675):1301--1305, 2004.

\bibitem{gent74}
A.~N. Gent.
\newblock Rubber and rubber elasticity: a review.
\newblock {\em J. Polymer Sci.}, 48(1):1--17, 1974.

\bibitem{gent59}
A.~N. Gent and P.~B. Lindley.
\newblock Internal rupture of bonded rubber cylinders in tension.
\newblock {\em Proc. Royal Soc. London A}, 249(1257):195--205, 1959.

\bibitem{gent91}
A.~N. Gent and C.~Wang.
\newblock Fracture mechanics and cavitation in rubber-like solids.
\newblock {\em J. Mater. Sci.}, 26(12):3392--3395, 1991.

\bibitem{gibb06}
Josiah~Willard Gibbs.
\newblock {\em The scientific papers of J. Willard Gibbs}, volume~1.
\newblock Longmans, Green and Company, 1906.

\bibitem{hein05}
M.~Heinemann, H.~Meinberg, J.~B{\"u}chs, H.-J. Ko{\ss}, and M.~B.
  Ansorge-Schumacher.
\newblock Method for quantitative determination of spatial polymer distribution
  in alginate beads using raman spectroscopy.
\newblock {\em Appl. Spectroscopy}, 59(3):280--285, 2005.

\bibitem{herz07}
E.~M. Herzig, K.~A. White, A.~B. Schofield, W.~C.~K. Poon, and P.~S. Clegg.
\newblock Bicontinuous emulsions stabilized solely by colloidal particles.
\newblock {\em Nature Mater.}, 6(12):966--971, 2007.

\bibitem{holt97}
J.~H. Holtz and S.~A. Asher.
\newblock Polymerized colloidal crystal hydrogel films as intelligent chemical
  sensing materials.
\newblock {\em Nature}, 389(6653):829, 1997.

\bibitem{hutc16}
S.~B. Hutchens, S.~Fakhouri, and A.~J. Crosby.
\newblock Elastic cavitation and fracture via injection.
\newblock {\em Soft matter}, 12(9):2557--2566, 2016.

\bibitem{hyma14}
A.~A. Hyman, C.~A. Weber, and F.~J{\"u}licher.
\newblock Liquid-liquid phase separation in biology.
\newblock {\em Ann. Rev. Cell Dev. Bio.}, 30:39--58, 2014.

\bibitem{jodd16}
B.~Joddar, E.~Garcia, A.~Casas, and C.~M. Stewart.
\newblock Development of functionalized multi-walled carbon-nanotube-based
  alginate hydrogels for enabling biomimetic technologies.
\newblock {\em Scientific reports}, 6, 2016.

\bibitem{koll13}
M.~Kolle, A.~Lethbridge, M.~Kreysing, J.~J. Baumberg, J.~Aizenberg, and
  P.~Vukusic.
\newblock Bio-inspired band-gap tunable elastic optical multilayer fibers.
\newblock {\em Adv. Mater.}, 25(15):2239--2245, 2013.

\bibitem{kund09}
Santanu Kundu and Alfred~J Crosby.
\newblock Cavitation and fracture behavior of polyacrylamide hydrogels.
\newblock {\em Soft Matter}, 5(20):3963--3968, 2009.

\bibitem{lame50}
V.~K. LaMer and R.~H. Dinegar.
\newblock Theory, production and mechanism of formation of monodispersed
  hydrosols.
\newblock {\em J. Am. Chem. Soc.}, 72(11):4847--4854, 1950.

\bibitem{lepe14}
E.~Lepeltier, C.~Bourgaux, and P.~Couvreur.
\newblock Nanoprecipitation and the ?ouzo effect?: Application to drug
  delivery devices.
\newblock {\em Adv. Drug Delivery Rev.}, 71:86--97, 2014.

\bibitem{macm15}
C.~W. MacMinn, E.~R. Dufresne, and J.~S. Wettlaufer.
\newblock Fluid-driven deformation of a soft granular material.
\newblock {\em Phys. Rev. X}, 5(1):011020, 2015.

\bibitem{mery56}
H.~T. Meryman.
\newblock Mechanics of freezing in living cells and tissues.
\newblock {\em Science}, 124(3221):515--521, 1956.

\bibitem{mitr16}
D.~M. Mitrea and R.~W. Kriwacki.
\newblock Phase separation in biology; functional organization of a higher
  order.
\newblock {\em Cell Commun. Signal.}, 14(1):1, 2016.

\bibitem{moll15}
A.~Molliex, J.~Temirov, J.~Lee, M.~Coughlin, A.~P. Kanagaraj, H.~J. Kim,
  T.~Mittag, and J.~P. Taylor.
\newblock Phase separation by low complexity domains promotes stress granule
  assembly and drives pathological fibrillization.
\newblock {\em Cell}, 163(1):123--133, 2015.

\bibitem{orak07}
N.~Orakdogen and O.~Okay.
\newblock Influence of the initiator system on the spatial inhomogeneity in
  acrylamide-based hydrogels.
\newblock {\em J. Appl. Polymer Sci.}, 103(5):3228--3237, 2007.

\bibitem{pate15}
A.~Patel, H.~O. Lee, L.~Jawerth, S.~Maharana, M.~Jahnel, M.~Y. Hein,
  S.~Stoynov, J.~Mahamid, S.~Saha, and T.~M. et~al. Franzmann.
\newblock A liquid-to-solid phase transition of the als protein fus accelerated
  by disease mutation.
\newblock {\em Cell}, 162(5):1066--1077, 2015.

\bibitem{riba17}
J.~A. Riback, C.~D. Katanski, J.~L. Kear-Scott, E.~V. Pilipenko, A.~E. Rojek,
  T.~R. Sosnick, and D.~A. Drummond.
\newblock Stress-triggered phase separation is an adaptive, evolutionarily
  tuned response.
\newblock {\em Cell}, 168(6):1028--1040, 2017.

\bibitem{schn00}
M.~J. Schnitzer, K.~Visscher, and S.~M. Block.
\newblock Force production by single kinesin motors.
\newblock {\em Nature Cell Bio.}, 2(10):718, 2000.

\bibitem{styl15}
Robert~W Style, Rostislav Boltyanskiy, Benjamin Allen, Katharine~E Jensen,
  Henry~P Foote, John~S Wettlaufer, and Eric~R Dufresne.
\newblock Stiffening solids with liquid inclusions.
\newblock {\em Nature Phys.}, 11(1):82--87, 2015.

\bibitem{tan16}
H.~Tan, C.~Diddens, P.~Lv, J.~G.~M. Kuerten, X.~Zhang, and D.~Lohse.
\newblock Evaporation-triggered microdroplet nucleation and the four life
  phases of an evaporating ouzo drop.
\newblock {\em Proc. Nat. Acad. Sci.}, 113(31):8642--8647, 2016.

\bibitem{vita03}
S.~A. Vitale and J.~L. Katz.
\newblock Liquid droplet dispersions formed by homogeneous liquid- liquid
  nucleation:?the ouzo effect?
\newblock {\em Langmuir}, 19(10):4105--4110, 2003.

\bibitem{wen11}
Q.~Wen and P.~A. Janmey.
\newblock Polymer physics of the cytoskeleton.
\newblock {\em Curr. Op. Solid State Mater. Sci.}, 15(5):177--182, 2011.

\bibitem{whee08}
T.~D. Wheeler and A.~D. Stroock.
\newblock The transpiration of water at negative pressures in a synthetic tree.
\newblock {\em Nature}, 455(7210):208, 2008.

\bibitem{zhu11}
J.~Zhu, T.~Li, S.~Cai, and Z.~Suo.
\newblock Snap-through expansion of a gas bubble in an elastomer.
\newblock {\em J. Adhesion}, 87(5):466--481, 2011.

\bibitem{zhu15}
L.~Zhu and C.~P. Brangwynne.
\newblock Nuclear bodies: the emerging biophysics of nucleoplasmic phases.
\newblock {\em Curr. Op. Cell Bio.}, 34:23--30, 2015.

\bibitem{zimb07}
J.~A. Zimberlin, N.~Sanabria-DeLong, G.~N. Tew, and A.~J. Crosby.
\newblock Cavitation rheology for soft materials.
\newblock {\em Soft Matter}, 3(6):763--767, 2007.

\bibitem{zwic14}
D.~Zwicker, M.~Decker, S.~Jaensch, A.~A. Hyman, and F.~J{\"u}licher.
\newblock Centrosomes are autocatalytic droplets of pericentriolar material
  organized by centrioles.
\newblock {\em Proc. Nat. Acad. Sci.}, 111(26):E2636--E2645, 2014.

\end{thebibliography}

\end{document}